\documentclass{aa}  

\usepackage{graphicx}
\usepackage{txfonts}

\usepackage[dvipsnames]{xcolor}
\usepackage{xspace}
\usepackage{anyfontsize}
\usepackage{comment}

\newcommand{\decode}{\textsc{decode}\xspace}
\begin{document}

   \title{Shedding light on the star formation rate-halo accretion rate connection and halo quenching mechanism via DECODE, the Discrete statistical sEmi-empiriCal mODEl}
   \titlerunning{\decode and galaxy quenching}
   \authorrunning{H. Fu et al.}


   \author{Hao Fu
            \inst{1,2},
            Lumen Boco\inst{3,4},
            Francesco Shankar\inst{2},
            Andrea Lapi\inst{4},
            Mohammadreza Ayromlou\inst{5,3},
            Daniel Roberts\inst{2},
            Yingjie Peng\inst{6,7},
            Aldo Rodr\'iguez-Puebla\inst{8},
            Feng Yuan\inst{1},
            Cressida Cleland\inst{9},
            Simona Mei\inst{9},
            Nicola Menci\inst{10}
          }

   \institute{Center for Astronomy and Astrophysics and Department of Physics, Fudan University, Shanghai 200438, People’s Republic of China\\
              \email{haofu@fudan.edu.cn}\and
            School of Physics and Astronomy, University of Southampton, Highfield, Southampton, SO17 1BJ, UK\and
            Universit{\"a}t Heidelberg, Zentrum f{\"u}r Astronomie, Institut f{\"u}r theoretische Astrophysik, Albert-Ueberle-Str. 2, 69120 Heidelberg, Germany\and
            SISSA, Via Bonomea 265, 34136 Trieste, Italy\and
            Argelander-Institut f\"ur Astronomie, Auf dem H\"ugel 71, D-53121 Bonn, Germany\and
            Department of Astronomy, School of Physics, Peking University, 5 Yiheyuan Road, Beijing 100871, China\and
            Kavli Institute for Astronomy and Astrophysics, Peking University, 5 Yiheyuan Road, Beijing 100871, China\and
            Universidad Nacional Aut\'onoma de M\'exico, Instituto de Astronom\'ia, A.P. 70-264, 04510, CDMX, M\'exico\and
            Université Paris Cité, CNRS(/IN2P3), Astroparticule et Cosmologie, F-75013 Paris, France\and
            INAF – Osservatorio Astronomico di Roma, via di Frascati 33, I-00078 Monte Porzio Catone, Italy
             }

   \date{Received September 15, 1996; accepted March 16, 1997}


%
 
  \abstract
   {}
   {The relative roles of the physical mechanisms involved in quenching galaxy star formation are still unclear. We tackle this fundamental problem with our cosmological semi-empirical model \decode (Discrete statistical sEmi-empiriCal mODEl), designed to predict galaxy stellar mass assembly histories, from minimal input assumptions. }
   {Specifically, in this work the star formation history of each galaxy is calculated along its progenitor dark matter halo by assigning at each redshift a star formation rate extracted from a monotonic star formation rate-halo accretion rate (SFR-HAR) relation derived from abundance matching between the (observed) SFR function and the (numerically predicted) HAR function, a relation that is also predicted by the TNG100 simulation. SFRs are integrated across cosmic time to build up the mass of galaxies, which may halt their star formation following input physical quenching recipes.  }
   {In this work we test the popular halo quenching scenario and we find that 1) the assumption of a monotonic relation between the SFR and HAR allows us to reproduce the number densities of the bulk of star-forming galaxies in the local Universe; 2) the halo quenching is sufficient to reproduce the statistics of the quenched galaxies and flat (steep) high-mass end of the stellar mass-halo mass relation (or SMF); and 3) to align with the observed steep (flat) low-mass end of the stellar mass-halo mass (or SMF) additional quenching processes in the least massive haloes are needed.}
   {\decode is an invaluable tool and will pave the way to investigate the origin of newly observed high-redshift objects from the latest ongoing facilities such as JWST and Euclid.}


   \maketitle
%

\section{Introduction}

Galaxies are among the most fascinating and mysterious objects in the Universe. They are believed to form and grow inside host dark matter haloes. However, while the evolution of dark matter haloes is somewhat well-established by cosmological N-body simulations, the evolution of the baryonic matter inside them is still hotly debated and subject to many uncertainties. One of the most intriguing facts is that while haloes evolve hierarchically, with larger structures being formed later by mergers of smaller ones, galaxies seem to follow a contrasting trend called downsizing. Observations and simulations suggest that massive galaxies are formed earlier in a burst of star formation and smaller galaxies formed over longer timescales, showing different trends even for galaxies of the same stellar mass \citep[e.g.][]{cowie_1996, thomas_2010, martin_navarro_2018a, baker_2024}. Explaining these multiple, and sometimes opposite, behaviours from a theoretical point of view is complex as they involve a deep understanding of the processes regulating star formation and quenching, i.e., the halting or significant reduction of star formation in some galaxies.

Several observations have shown the existence of two main categories of galaxies: 1) star-forming and 2) quiescent or quenched. Star-forming galaxies are galaxies that are actively forming stars; they are bluer, less massive and younger on average. Quenched galaxies have low or no ongoing star formation, and are redder, more massive and older on average \citep[][]{blanton_2003, baldry_2004, wuyts_2011, wetzel_2012, van_der_wel2014, dimauro_2019}. The origins of these differences are rooted in the physical mechanisms responsible for quenching, which is still highly debated in the literature, with many works advocating different reasons for the shutdown of star formation \citep[][]{granato_2004, dekel_2006, martig_2009, lapi_2018, gensior_2020}.

Quenching scenarios are typically classified into two categories: 1) internal quenching and 2) environmental quenching. Internal quenching, or mass quenching \citep[][]{peng_2010, peng_2012} includes all those mechanisms related to internal galactic processes, such as gas heating and outflow due to stellar winds, supernova feedback \citep[e.g.][]{larson_1974, dekel_1986}, active galactic nucleus (AGN) feedback from the central supermassive black hole (SMBH) \citep[e.g.][]{silk_1998, granato_2004, bower_2006, croton_2006, somerville_2008, fabian_2012, fang_2013, cicone_2014, lapi_2018}, mergers \citep[e.g.][]{schawinski_2010}, morphological quenching \citep[e.g.][]{martig_2009} and halo quenching, where the hot gas of the galaxy is prevented from cooling when the mass of the host halo surpasses a certain threshold, due to the shock heating that heats the interstellar medium \citep[][]{birnboim_2003, dekel_2006}. Environmental quenching includes several distinct mechanisms due to the interaction between the galaxy and the environment. Renowned quenching mechanisms, such as ram-pressure stripping, starvation and strangulation, fall in this category \citep[e.g.][]{larson_1980, gunn_1972, cowie_1977}.

Many works have attempted to model galaxy quenching. Semi-analytical models use parametric recipes for implementing the physical processes involved in quenching \citep[e.g.][]{cattaneo_2020, ayromlou_2021, koutsouridou_2022, lagos_2023}. Hydrodynamical simulations combine subgrid prescriptions with the solution of the equations for feedback physics \citep[e.g.][]{donnari_2021a, donnari_2021b, gandhali_2021, piotrowska_2022}. Despite this great effort, different models produce quite different ranges of predictions, where the number and redshift of quiescent galaxies vary from model to model. For this reason, the primary physical mechanism leading to quenching, as well as the relation among quenching, mass, and galaxy morphology, are still not fully understood.

In this paper we present a new semi-empirical model, \decode, an updated version of the Discrete statistical sEmi-empiriCal mODEl presented in \citet{fu_2022, fu_2024}. \decode is a semi-empirical model originally designed to explore, in a fully data-driven fashion, the connection between star formation histories and merger rates of galaxies for a given input stellar mass-halo mass (SMHM) relation. By marginalizing over the specifics of the complex baryonic physics, \decode's main goal is to pin down the star formation histories of galaxies of different stellar mass in a robust manner, only guided by the data and an assumed underlying dark matter assembly history. Although it is less ambitious than a numerical simulation or a semi-analytic model, \decode relies on a very limited set of assumptions and parameters, making it an extremely transparent and model-independent framework to test various quenching mechanisms. Given the data-driven nature, \decode departs from more traditional ab initio approaches, as it requires in input some data sets in order to predict some specific outputs, in a flexible and transparent way, that are then tested against other independent data sets. Our approach is thus particularly useful to highlight possible inconsistencies in different data sets and to test some basic assumptions, for example on the process of halo quenching, as carried out in this work.

In this work we further upgraded \decode to follow the evolution of galaxy assembly in each single dark matter halo extracted from a cosmological box. In the previous work, we followed mean growth histories of galaxies by making use of the (mean) SMHM relation to assign at each epoch galaxies to the main progenitor halo. Here instead we enabled \decode to predict the mass growth via accretion and mergers of each galaxy by making use of the abundance matching between star formation rates and dark matter halo accretion rates. The reason behind the choice of this new type of abundance matching is twofold. On the one hand, quenching mechanisms decouple the co-evolution of galaxies and dark matter haloes on the SMHM relation (e.g. \citealt{peng_2012, man_2019, lyu_2023}), making the abundance matching between stellar mass and halo mass less robust, especially at the bright end. On the other hand, the connection between mass accretion rates provides a much more flexible and straightforward way to test the quenching mechanisms without degeneracies with many others parameters of the model.

We made use of the upgraded \decode to specifically focus on the role of the halo quenching due to shock heating, which is one of the possibly invoked mechanisms responsible for the halting of star formation in massive central galaxies at least at $z\lesssim 2$. We  show how this mechanism regulates the star formation activity in galaxies, by analysing the predicted mass growth histories, stellar mass functions (SMFs), fractions of quenched galaxies, and SMHM relation as a function of cosmic time compared to current available data. In the next papers of this series we will investigate the role of various other quenching mechanisms such as the black hole feedback, mergers, morphological and environmental quenching, with the final goal of unveiling the most effective physical processes that, in fully data-driven cosmological setting, are able to reproduce current data sets.

This paper is structured as follows. 
In Sect. \ref{sec_decode_implementation} we define the main lines of our methodology, and we describe the abundance matching and the way we grow our galaxies. In Sect. \ref{sec_tng_test} we test our input assumption on the abundance matching between galaxy star formation rate and dark matter halo accretion rate. In Sect. \ref{sec_results} we show our results on the galaxy abundances, fraction of quenched and SMHM relation. Finally, in Sects. \ref{sec_discuss} and \ref{sec_conclu} we discuss our findings and draw our conclusions. In this paper we adopt the $\Lambda$CDM cosmology with best fit parameters from \citet{planck2018_cosmo_params} (i.e. $(\Omega_{\rm m}, \Omega_\Lambda, \Omega_{\rm b}, h, n_{\rm S}, \sigma_8) = (0.31, 0.69, 0.049, 0.68, 0.97, 0.81)$); all the input and reference data sets adopted here use a \citet{chabrier_2003} stellar initial mass function.

\begin{figure*}
    \includegraphics[width=\textwidth]{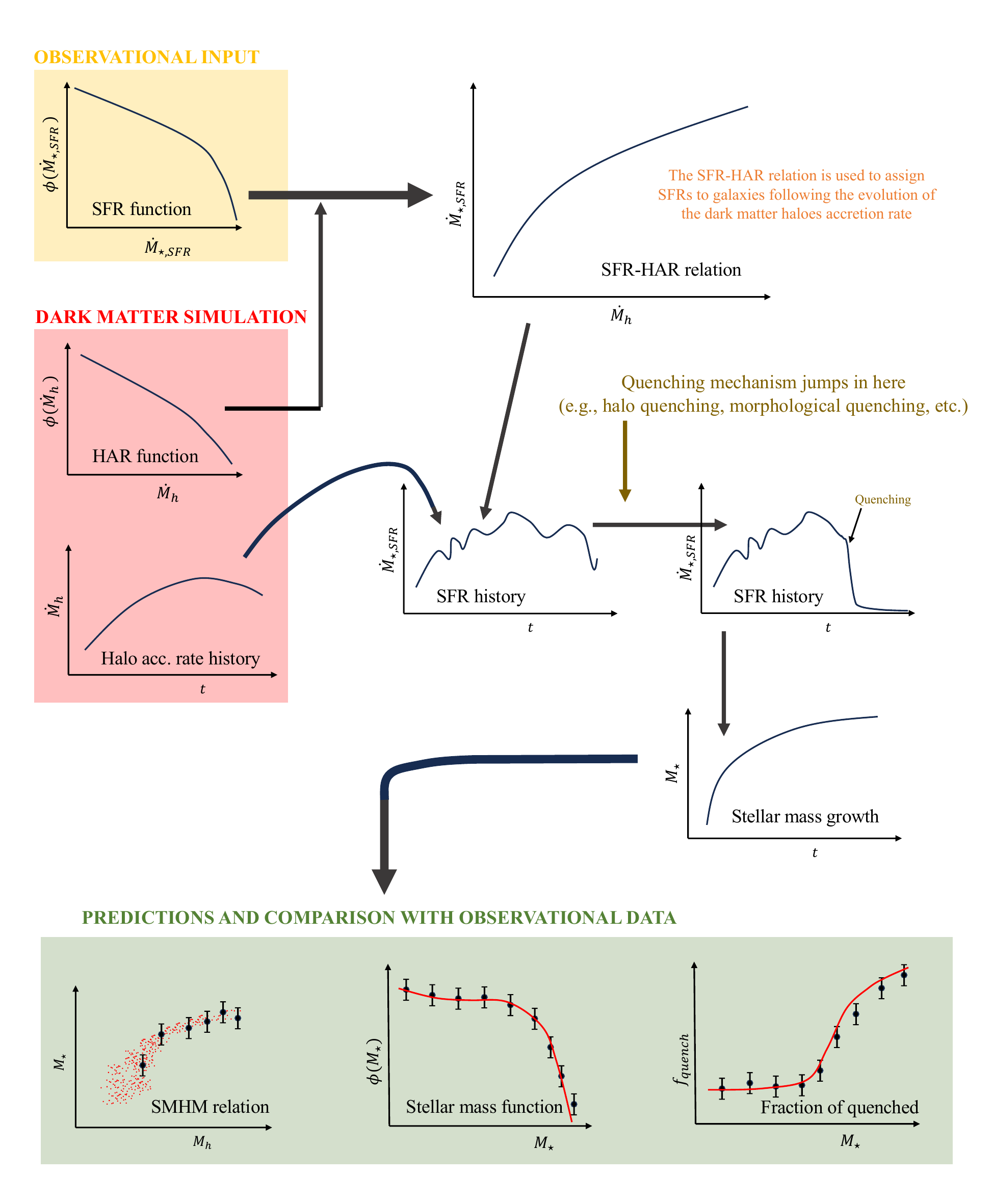}
    \caption{Cartoon showing the methodology used to form and evolve galaxies in \decode. The SFR-HAR relation at each redshift is computed via abundance matching between the observed SFR function and HAR function from simulations. SFRs are assigned to galaxies following the accretion rate history of their dark matter haloes. We drop the SFR instantaneously when the galaxy is quenched. Galaxy stellar mass growths are constructed by integrating the SFR and quenching the star formation via the mechanism we are testing. Finally, we make our predictions on the SMHM relation, SMF, and quiescent fraction, and compare them with the observations.}
    \label{fg_cartoon_decode2}
\end{figure*}

\section{The DECODE implementation}\label{sec_decode_implementation}

In this Section, we describe \decode and how we use it to evolve galaxies within their host dark matter haloes. \decode relies on merger trees generated analytically, and halo accretion histories are converted into galaxy stellar mass assemblies via a scaling relation between the galaxy star formation rate (SFR) and dark matter halo accretion rate (HAR), called the SFR-HAR relation. On top of this semi-empirical baseline, one can test any quenching scenario, which will generate a diverse set of outputs. In this first paper of the series, we test the halo quenching scenario, whose outputs are compared with up-to-date observational data. In future work, we will investigate the role of AGN feedback and morphological transformations in shaping the galaxy star formation histories.

The main steps of the methodology are summarised as follows:
\begin{itemize}
    \item computing the SFR-HAR relation;
    \item assigning SFRs to galaxies;
    \item building galaxy stellar mass growth histories;
    \item quenching the star formation.
\end{itemize}

Figure \ref{fg_cartoon_decode2} shows a schematic of \decode. In brief, we first compute the star formation rate-halo accretion rate (SFR-HAR) relation via abundance matching between the dark matter HAR function and the SFR function; the latter is derived from observed luminosity functions. Secondly, galaxies are assigned with a SFR as a function of redshift via the SFR-HAR relation following the mass accretion history of their host haloes, and their star formation is quenched following the tested quenching recipe. Finally, we make our predictions for the galaxy SMF, quenched fraction and SMHM relation, and we compare them with the same quantities derived from independent and distinct data sets.

In the build-up of the stellar mass growths, \decode is inspired by various semi-empirical models in the literature (e.g. \citealt{moster_2018, behroozi_2019}). 
However, the approach presented here is significantly distinct from previous attempts in two main respects. Firstly, our method relies on the direct connection between the galaxy SFR and the HAR (and not, e.g. halo circular velocity), which is a physically grounded assumption and indeed also seen in hydrodynamical simulations, as discussed in Sect. \ref{sec_tng_test}. Secondly, the SFR-HAR relation is derived from abundance matching techniques at any given redshift, thus bypassing any heavy fitting and minimizing the input assumptions and parameters. Our methodology is closer to the one devised in the semi-empirical model TOPSEM, presented in \citet{boco_2023}, although it still differs from it in some crucial aspects. \decode avoids the initialization of galaxies through an input SMF at redshift $z=0$ as it builds up the stellar mass directly integrating the star formation rate across time rather than using the specific SFR. Moreover, in the present rendition of \decode, quenched galaxies are not empirically assigned as input in the model, as in TOPSEM, but they are progressively generated through time via some physically motivated recipe (Fig. \ref{fg_cartoon_decode2}) which can then be tested against observations. In this work we assume star formation is halted uniquely by halo quenching.

    \subsection{From dark matter haloes to galaxies}\label{sec_method_DM_cat}

    We started at a given redshift of interest $z$ creating a catalogue of $N=10^5$ parent dark matter haloes with mass $M_{\rm h,par} > 10^{11}\, M_\odot$, equivalent to a box of 180 Mpc on a side, randomly extracted from the halo mass function (HMF) from \citet{tinker_2008} (similarly to what was done in \citealt{fu_2022}, \citealt{fu_2024}, and \citealt{boco_2023}). This choice represents a good balance between mass resolution, computational efficiency and statistical significance of the sample for massive haloes. 
    We computed the mass accretion history of each dark matter halo in the box using the SatGen code presented in \citet{jiang_2021}, which is based on the \citet{parkinson_2008} algorithm and well reproduces the mean mass accretion and halo mass variance of dark matter haloes of the MultiDark Planck simulation up to redshift $z\sim 9$ with mass resolution down to $M_{\rm h} \sim 10^9 \, M_\odot$. 
    To each dark matter halo we also assigned the masses and redshifts of infall of its first-order satellites, merging with the main progenitor. As discussed in \citet{fu_2022}, the first-order satellites are sufficient to well represent the full merger history undergone by the central galaxy, composing $\gtrsim85\%$ of the number of merging satellites at any halo mass (as shown in Sect. 4.2 of \citealt{fu_2022}). Central galaxies then self-consistently grow in stellar mass via direct integration of their SFRs along the main progenitors and additional mergers with galactic satellites. We repeated the above procedure at any given redshift of interest and build the SMF as the sum of the central galaxies and (surviving) satellites surviving down to that redshift. 

    \subsection{Abundance matching}\label{sec_method_AM_sfr_har}

    The key element of \decode is the mapping between the galaxy SFR and host halo accretion rate, the SFR-HAR relation. We computed the latter via the abundance matching between observed SFR distributions and the HAR distributions following the same formalism put forward in \citet[][Eq. 37 therein]{aversa_2015}, through which we calculate the mean SFR at given HAR using the equation
    \begin{equation}\label{eq_aversa_AM}
    \begin{split}
        \int_{\log \dot{M}_{\rm \star, SFR}}^{+\infty} \phi(\dot{M}_{\rm \star, SFR}', z) \mathrm d \log \dot{M}_{\rm \star, SFR}' = \;\;\,\, &\\ 
        \int_{-\infty}^{+\infty} \frac{1}{2} \mathrm{erfc} \Bigg\{ \frac{\log \dot{M}_{\rm h} (\dot{M}_{\rm \star, SFR}) - \log \dot{M}_{\rm h}' }{\sqrt{2} \Tilde{\sigma}_{\log \dot{M}_{\rm \star, SFR}}} \Bigg\} & \cdot \phi( \dot{M}_{\rm h}', z) \mathrm d \log \dot{M}_{\rm h}' \; ,
    \end{split}
    \end{equation}
    where $\Tilde{\sigma}_{\log \dot{M}_{\rm \star,SFR}} = \sigma_{\log \dot{M}_{\rm \star,SFR}} / \mu $. Here, $\sigma_{\log \dot{M}_{\rm \star,SFR}}$ is the Gaussian scatter in star formation rate at fixed halo accretion rate, and $\mu = \mathrm d \log \dot{M}_{\rm \star,SFR} / \mathrm d \log \dot{M}_{\rm h}$ is the derivative of the star formation rate with respect to the halo accretion rate. This recipe provides a fast and flexible tool to compute the SFR-HAR relation numerically with only two ingredients, namely the HAR and SFR distributions, and one input parameter, namely the scatter in $\log_{10} ({\rm SFR})$ at fixed $\log_{10} ({\rm HAR})$, without any pre-defined analytic fitting formula that requires a heavy Markov chain Monte Carlo (MCMC) exploration and the introduction of free parameters in the model. The SFR-HAR relation computed from Eq. (\ref{eq_aversa_AM}) was then used to grow galaxies within dark matter haloes. We chose a scatter in SFR at fixed HAR of $0.4$ dex, suggested by the results of the TNG simulation, even though we found that \decode's stellar mass growths do not change appreciably by changing this scatter within reasonable values ($0.3-0.5$ dex).

    \begin{figure}
        \centering
        \includegraphics[width=\hsize]{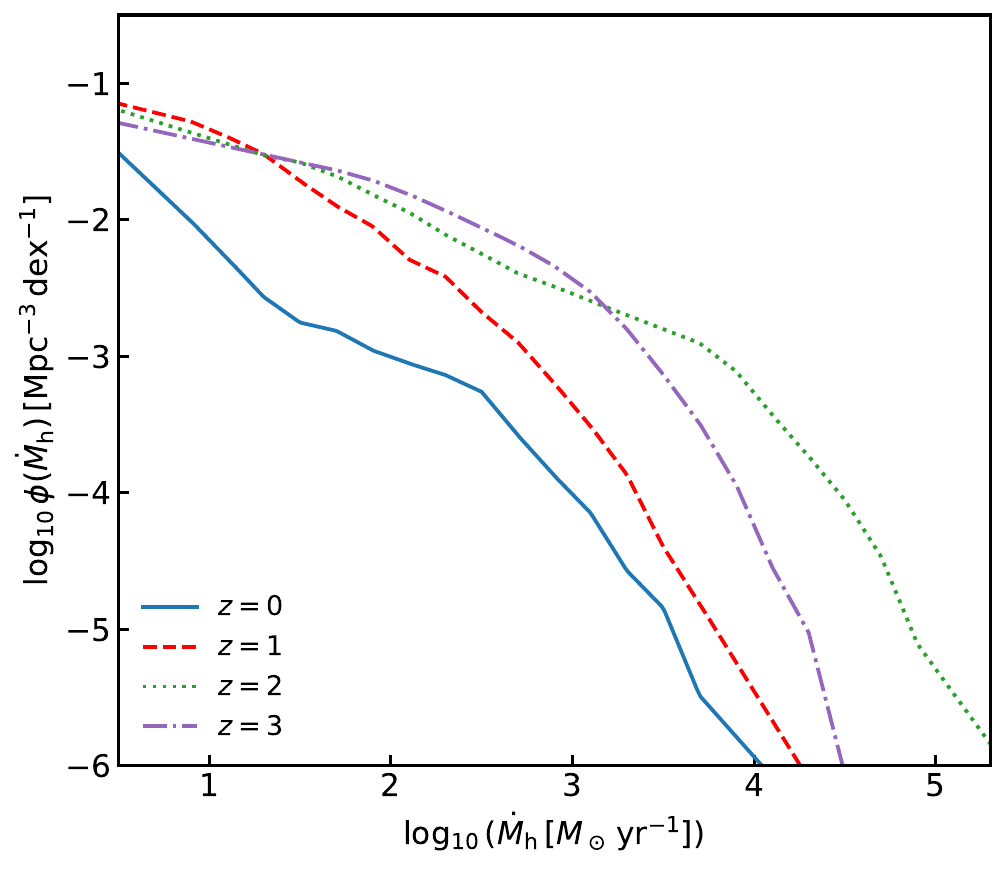}
        \includegraphics[width=\hsize]{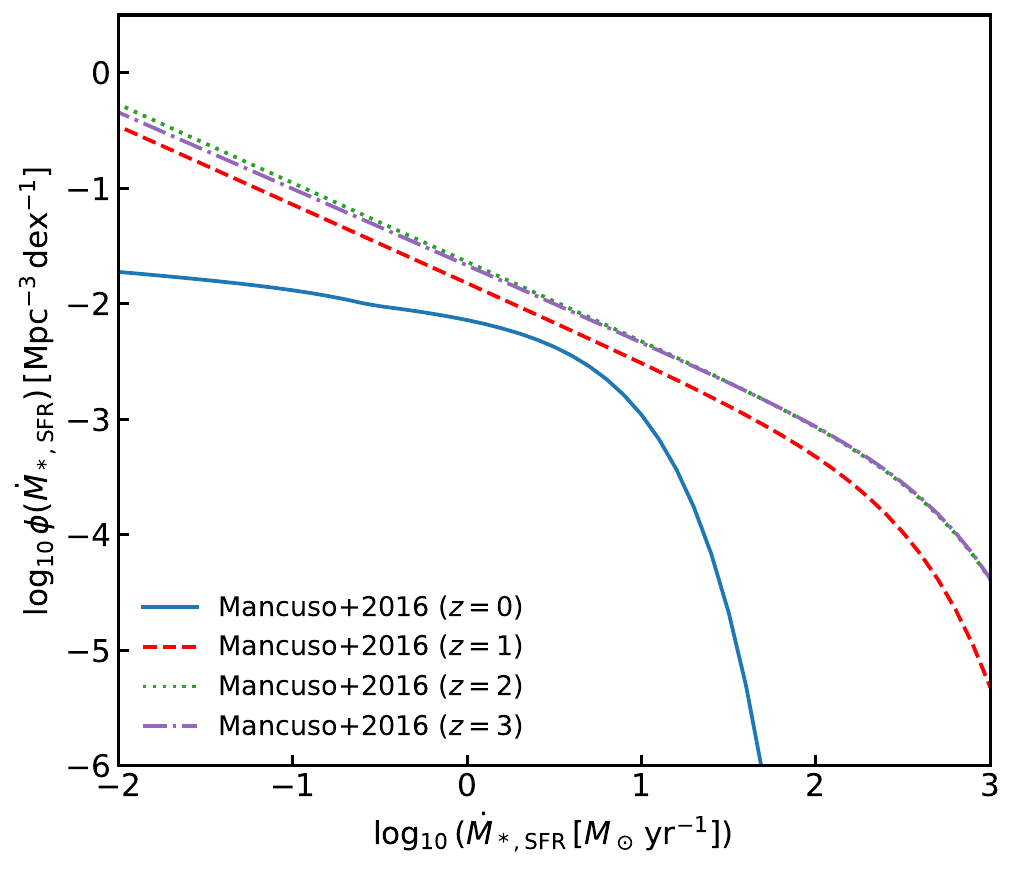}
        \caption{Upper panel: Halo accretion rate function at redshifts $z=0$, $1$, $2$ and $3$, for the halo quenching scenario where haloes hosting star-forming galaxies are removed by the mass threshold $M_{\rm h} \sim 10^{12}\, M_\odot$. Lower panel: Star formation rate function from \citet{mancuso_2016} in the same redshift bins, as labelled.}
        \label{fg_phi_sfr}
    \end{figure}

    The two main ingredients entering in Eq. (\ref{eq_aversa_AM}) are the observational star formation rate function and the theoretical halo accretion rate function. We made use of the HAR function sampled from the mock catalogue at any redshift, described in Sect. \ref{sec_method_DM_cat}. Given the focus on the halo quenching scenario in this paper, where galaxies within haloes with mass $M_{\rm h,lim} \gtrsim 10^{12} \, M_\odot$ are considered quiescent (as we describe in Sect. \ref{sec_method_haloquenching}), we removed all those haloes from the abundance matching. As for the SFR function, we made use of the determination of \citet{mancuso_2016} (see also \citealt{boco_2019, boco_2021}), who derived a fitting formula for the SFR function starting from the galaxy luminosity functions (LFs) and converted luminosity to SFR. Specifically, \citet{mancuso_2016} combines LF reconstructions in UV band, from HST and GALEX data, and in far-IR band, from the Herschel data, to be unbiased by dust obscuration at the bright end. The fit extends up to $z\lesssim 3$, given the sensitivity limit of far-IR surveys. However, the SFR function fit at $z>3$ was validated via a continuity equation approach against a number of independent observables: galaxy number counts at far-IR and radio wavelengths, the main sequence of star-forming galaxies, and the stellar mass function (\citealt{lapi_2017}). In the present paper we limit our analysis to $z<3$ to avoid any untested extrapolations of the \citet{mancuso_2016} fit to the SFR function. 
    Figure \ref{fg_phi_sfr} shows the HAR (upper panel) and SFR (lower panel) number densities from \citet{mancuso_2016} at different redshifts. The SFR function does not evolve much at redshifts $z\sim 1.5-3$ and drops quickly below $z\lesssim1$, both at the faint and the bright ends, possibly due to a more rapid increase in the fraction of quiescent galaxies as well as a general decrease in the SFR density of galaxies below $z\sim 1-2$ (e.g. \citealt{bouwens_2009, cucciati_2012, madau_2014, katsianis_2017, magnelli_2024}). In Appendix \ref{Appendix:additional_SFRs}, we show that the \citet{mancuso_2016} fits are aligned with other determinations of the SFR functions in the literature at $z<3$, and we verified that switching to any of these alternative estimates yields qualitatively similar results. 
    

    \subsection{Tracking stellar mass evolution in galaxies}\label{sec_method_grow_Mstar}

    \subsubsection{Central galaxies}

    The SFR-HAR relation generated via abundance matching was used to assign the SFR to galaxies following the accretion rate of the host dark matter haloes at each redshift. We then simply integrated the SFR forward across cosmic time to predict the stellar mass growth of each galaxy. Following the quenching mechanism that we aim to test (i.e. shock heating) we instantaneously halted the star formation, as detailed in Sect. \ref{sec_method_haloquenching}. 
    When computing the stellar mass, we corrected the SFR to include the loss rate of the stellar mass that goes into the interstellar medium using the \citet{reimers_1975} factor, $1-\mathcal{R}$, with $\mathcal{R}=0.4$ (e.g. \citealt{renzini_1988, girardi_2000, pietrinferni_2004}).\footnote{We have checked that employing the \citet{leitner_2011} recipe yields the same results.}

    At each redshift, we obtained a population of surviving\footnote{We use the term 'surviving' to define those centrals that have become satellites of another galaxy before a given redshift.} central galaxies, for which we were able to compute the relative amount of star-forming and quenched galaxies, according to the level of their SFRs. We stress that the fraction of quenched galaxies is an actual prediction of the model, which makes this work distinct from previous, more descriptive semi-empirical models, since it allows us to directly test the quenching physical mechanisms.

    \subsubsection{Satellite galaxies}\label{sec_method_sat_gals}

    An additional important process that cannot be neglected in the galaxy mass assembly, especially in more massive galaxies, is the merger with other galaxies (e.g. \citealt{guo_2008, oser_2010, cattaneo_2011, lackner_2012, lee_2013, pillepich_2014, rodriguez_gomez_2016, qu_2017, clauwens_2018, pillepich_2018_Mstar_content, monachesi_2019, davison_2020, grylls_2020b, fu_2022, fu_2024}). We accounted for the contribution of mergers to the stellar mass growth in each galaxy via the analytical merger tree generated via SatGen, as described in Sect. \ref{sec_method_DM_cat}, and the merging timescale following the recipe described in Sects. 3.4 and 3.6 of \citet{fu_2022}, who provided an accurate fit to the fudge factor to correct the dynamical timescales of dark matter subhaloes from numerical simulations (e.g. \citealt{boylan_kolchin_2008, mccavana_2012}). We then converted the dark matter merger histories into cumulative stellar mass merger histories using the SMHM relation predicted by the model itself for central galaxies. The stellar masses of the satellite galaxies at the centre of infalling dark matter subhaloes were initialised from the mean and scatter of the SMHM relation characterizing central galaxies at that epoch of infall. We note that simply initializing infalling subhaloes by randomly extracting from the population of parent haloes of similar mass as in, e.g. \citet{hopkins_2009} or \citet{shankar_2014}, yields very similar results. 

    Also when predicting the galaxy SMF one should take into account the satellite galaxies. While in the massive part of the SMF satellites are approximately negligible, in the fainter part below the knee ($M_\star \lesssim 10^{11}\, M_\odot$) they gradually become more significant (e.g. \citealt{behroozi_2010, grylls_2020a, fu_2022}). Therefore, it is important to have a good recipe to account for the satellite abundances before computing the total galaxy population, as for example the bright end of the SMF may be significantly underestimated in the case that the contribution of satellites is missing. 



    \begin{figure}
        \centering
        \includegraphics[width=\hsize]{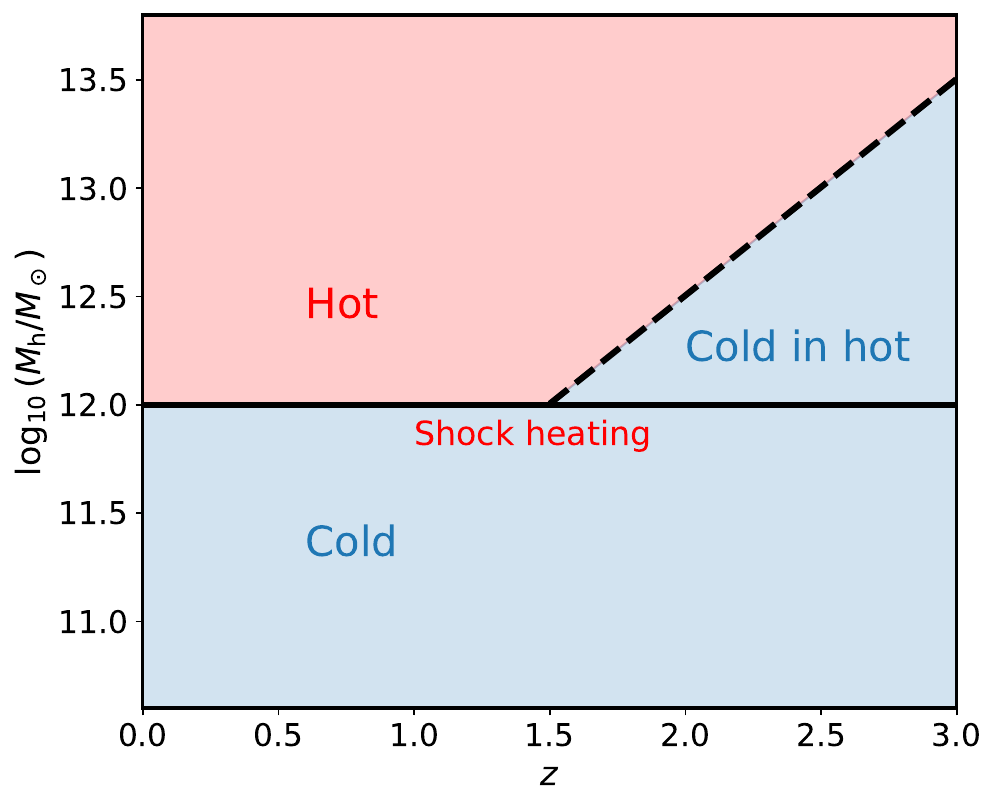}
        \caption{Schematic view of the halo quenching mass threshold. Galaxy living in haloes with mass above the threshold (black solid line) suppress their star formation due to shock heating. The area below the threshold and below the dashed line represents the region where cold streams can exist and still allow star formation.}
        \label{fg_scheme_shockheating}
    \end{figure}

    \begin{figure*}
        \includegraphics[width=\textwidth]{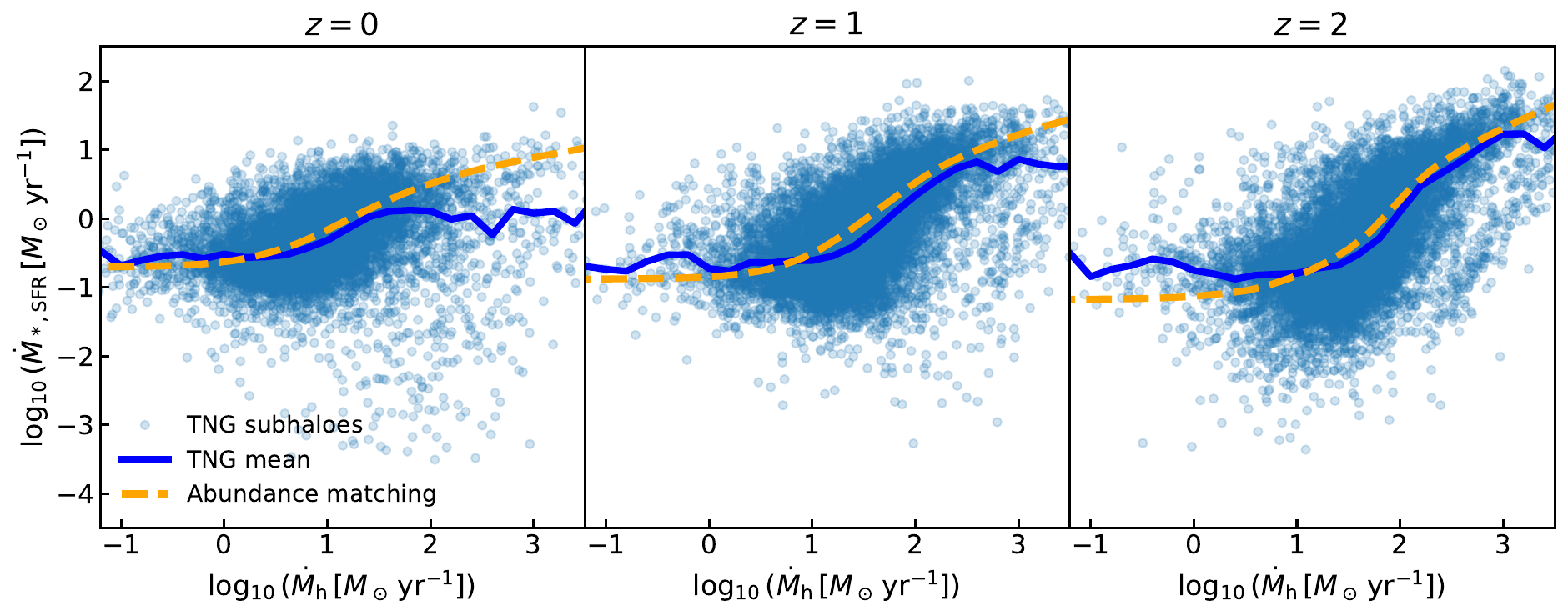}
        \caption{Distribution of the TNG central subhaloes/galaxies on the star formation rate-halo accretion rate plane (blue dots), and mean scaling relation from the TNG (blue lines) and computed using the TNG's inputs via the abundance matching described in Sect. \ref{sec_method_AM_sfr_har} (orange dashed lines) at redshifts $0$, $1$ and $2$.
        }
        \label{fg_SFR_HAR_TNG}
    \end{figure*}

    \subsection{Treatment of halo quenching}\label{sec_method_haloquenching}
    
    The main quenching mechanism that we consider in this paper is the halo quenching (e.g. \citealt{birnboim_2003, keres_2005, dekel_2006, dekel_2008}). The aforementioned works have shown that in massive haloes (e.g. massive galaxies, groups, lately formed clusters) the supersonic infalling gas becomes shock-heated at the virial temperature, heating the interstellar medium and preventing gas from cooling and triggering star formation. 
    On the other hand, below this threshold rapid cooling dominates over the shock gas pressure. \citet{birnboim_2003} and \citet{dekel_2006} computed the typical threshold halo mass (known as golden halo mass) as being equal to $M_{\rm h,lim} \sim 10^{12} \, M_\odot$.

    As described in Sect. \ref{sec_method_grow_Mstar}, for each galaxy in our catalogue, we assigned the SFR via the SFR-HAR relation computed via abundance matching, and we integrated the SFR across cosmic time to build up the stellar mass of each galaxy. 
    Then, we truncated the star formation when the dark matter halo mass exceeded the threshold value $M_{\rm h,lim}$ and we labelled the galaxy as quenched. Since physically not all galaxies quench when their host halo reaches exactly $M_{\rm h,lim} \sim 10^{12} \, M_\odot$, we included some dispersion around this quenching halo mass, for which we found a value of $0.4$ dex to best suit the outputs to the data. By increasing or decreasing this parameter the shape of the quenched fractions as a function of stellar mass would simply flatten or sharpen.

    Figure \ref{fg_scheme_shockheating} shows a schematic of the halo quenching threshold. All haloes above the threshold mass limit (black solid line) undergo shock heating that prevents gas from cooling (blue and red areas). \citet{dekel_2006} suggested that at $z\gtrsim 1.5$, 
    even in haloes above the canonical mass threshold of $M_{\rm h,lim} \sim 10^{12} \, M_\odot$ there is still room for cooling, at least below a certain redshift-dependent mass (see dashed line in Fig. \ref{fg_scheme_shockheating}). Only below the limit marked by the dashed and solid lines, cold streams could be formed, allowing disc growth and star formation. We checked that the effect of the inclusion of the cold-in-hot mode was minimal in our results. Therefore, we set our quenching threshold simply as that marked by the blue and red areas.

\section{Testing the SFR-HAR assumption}\label{sec_tng_test}

    Before applying \decode to predict the star formation histories and abundances of galaxies, we tested the input assumption on the SFR-HAR connection via the TNG simulation. Specifically, our aim was to test in a self-consistent model like TNG, whether the galaxy SFRs and host dark matter HARs are connected by a monotonically increasing relation, and to determine the impact of the inclusion of a time delay between the two quantities possibly due to the cooling time of the accreted gas. Moreover, we checked if, by starting from the SFR function and HAR function generated by the TNG simulation, the mean SFR-HAR relation of the TNG can be reproduced via our abundance matching technique described in Sect. \ref{sec_method_AM_sfr_har} using TNG's inputs.

    \subsection{The TNG simulation}

    In this test we made use of the data from the TNG100 simulation, a component of the IllustrisTNG project \citep[hereafter TNG;][]{nelson_2019, pillepich2018Simulating, springel2018first, marinacci2018first, naiman2018first}. The TNG simulations were performed using the moving-mesh \textsc{arepo} code \citep[][]{springel_2010}. The TNG employs subgrid modelling of galaxy formation-related processes such as gas cooling, star formation, stellar evolution, AGN feedback, and black hole processes (e.g. \citealt{pillepich2018Simulating, weinberger_2017}). The TNG simulations encompass different volume box sizes and mass resolutions. Here, we utilised the TNG100 simulation, which was run on a 100 Mpc box. 

    In order to test our assumptions 
    of monotonicity between SFR and HAR and whether this relation can be recovered via a direct abundance matching technique, we retraced each step of our methodology described above but directly applied to the TNG outputs. More specifically, we first 
    traced each halo and its central galaxy back in time to their first appearance in the simulation. This was done using subhalo merger trees produced by the SubLink algorithm, which constructs merger trees at the subhalo level \citep[][]{Rodriguez-Gomez2015}. We then measured the changes in stellar mass and halo mass for each galaxy and halo over time, computing the SFR and HAR for each central galaxy and its host dark matter halo, between each snapshot and the previous one (corresponding to a $\sim 100$ Myr step in cosmic time).
    
    \subsection{The SFR-HAR connection}

    \begin{figure}
        \includegraphics[width=\hsize]{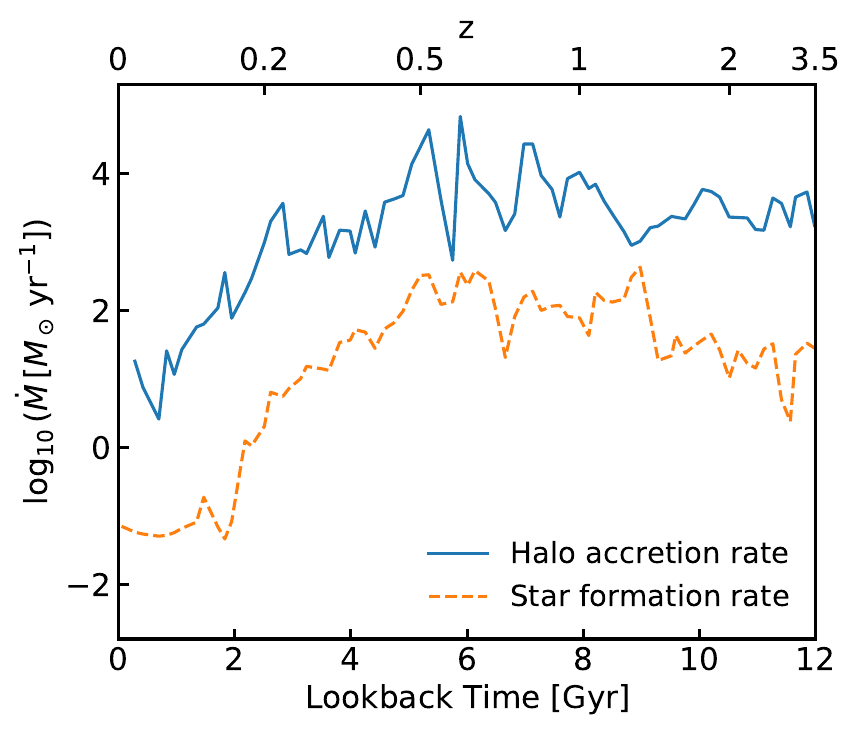}
        \caption{Example of evolution of the star formation rate and halo accretion rate of one central galaxy of stellar mass $M_\star \sim 10^{11.9}\, M_\odot$ at $z=0$ from the TNG simulation.}
        \label{fg_example_gal_evo_TNG}
    \end{figure}

    Figure \ref{fg_SFR_HAR_TNG} shows the SFR-HAR relation at different redshifts as computed from the TNG simulation. The plot shows that galaxy SFRs and host halo accretion rates appear connected via a monotonically increasing relationship and, as checked, with a symmetric dispersion around the mean, which supports our assumption of a Gaussian dispersion in the abundance matching algorithm. Moreover, when taking the TNG's SFR function, HAR function and scatter as input, \decode's abundance matching is able to well reproduce the mean SFR-HAR relation of the simulation at all redshifts. 
    The SFR and HAR functions in the TNG are computed by sampling the SFRs and HARs of the subhaloes in the simulation hosting star-forming galaxies, distinguished via the cut in specific SFR $\dot{M}_{\rm \star,SFR} /M_\star \lesssim 10^{-11}\, {\rm yr}^{-1}$ without any further assumption. We note that the conclusions on the robustness of the abundance matching do not alter since it is merely a self-consistency test. We also checked that by employing the TNG's SFR-HAR relation in \decode's framework, as expected by design, we can reproduce TNG's predicted star formation histories in star-forming galaxies at $z\lesssim 3$ using \decode's methodology of integrating the SFRs assigned via abundance matching forward in time. 
    
    State-of-the-art hydrodynamic simulations such as TNG predict a close correlation between the SFR and HAR. Figure \ref{fg_example_gal_evo_TNG} shows an example of co-evolution of the SFR of a galaxy of stellar mass $M_\star \sim 10^{11.9}\, M_\odot$ at $z=0$ and the accretion rate of its host dark matter halo versus lookback time in the TNG simulation. The trend shows that an overall increase (or decrease) in halo accretion rate is mirrored by an increase (or decrease) in SFR, and that there is no apparent time delay between SFR and HAR. We checked that this behaviour is followed by all galaxies in the TNG with any stellar mass at the present day, further supporting the robustness of \decode's methodology. We stress that the HARs and SFRs in the TNG simulation were computed over a time binning of $\sim 100$ Myr, whereas the gas cooling timescales could be shorter and its effect washed out due to resolution effects. Irrespective of this, we checked that adding a delay of 100 Myr between SFR and HAR, when applying the abundance matching procedure, does not significantly alter the predicted SFR-HAR relation.

\section{Results}\label{sec_results}

In this Section, we present \decode's predictions for the galaxy stellar mass assembly when adopting as input the SFR-HAR relation and employing the halo quenching as dominant quenching process for the massive galaxies below $z\lesssim2$. In particular, we show the output stellar mass-halo mass (SMHM) relations, stellar mass function (SMF), and fraction of quenched galaxies, and discuss how \decode's predictions compare to the latest observational determinations in the literature.

    \subsection{The SFR-HAR relation}\label{sec_sfr_har_relation}

    \begin{figure}
        \centering
        \includegraphics[width=\hsize]{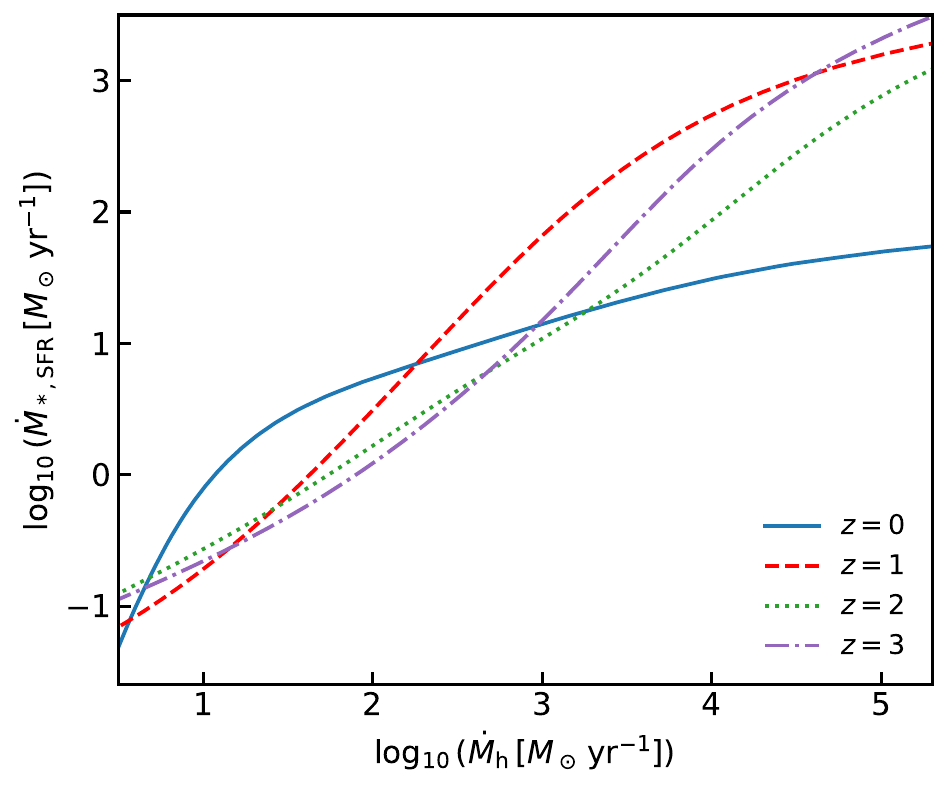}
        \caption{Star formation rate-halo accretion rate relation at redshifts $z=0$, $1$, $2$ and $3$, from the abundance matching using as input the star formation rate function from \citet{mancuso_2016}.}
        \label{fg_sfr_har}
    \end{figure}

    \begin{figure}
        \centering
        \includegraphics[width=\hsize]{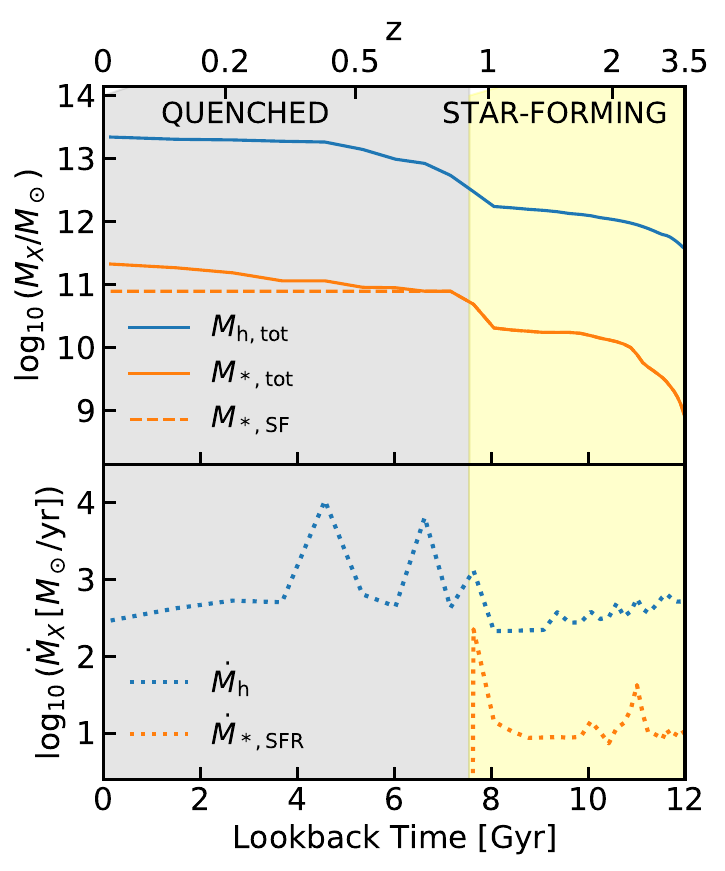}
        \caption{Example of evolution of a galaxy of stellar mass $M_\star \sim 10^{11.5}\, M_\odot$ from the catalogue, for the case of halo quenching. The upper and lower panels show the evolution in mass and growth rate, respectively. The blue solid and dotted lines show the halo mass assembly and accretion rate history, respectively. The orange solid and dotted lines show the stellar mass growth and star formation rate, respectively. Additionally the orange dashed line shows the case of star formation-only without mergers. At redshift $z\sim0.9$, when the halo mass reaches the threshold value, the star formation rate drops and the galaxy is quenched.}
        \label{fg_mass_acc}
    \end{figure}

    Figure \ref{fg_sfr_har} shows the SFR-HAR relation at $z=0$, $1$, $2$, and $3$ computed from abundance matching using the \citet{mancuso_2016} SFR function and the HAR function described in Sect. \ref{sec_decode_implementation}. In particular, at redshifts $z\gtrsim1$ the relation is well represented by a log-linear function with constant slope of $\sim 1.5$, while towards lower redshifts the SFR-HAR relation is characterised by a bending at high HARs. 
    These SFR-HAR relations are then used to assign SFRs to galaxies. 
    As described in Sect. \ref{sec_decode_implementation}, the stellar mass assembly histories are computed by assigning SFRs following the HAR histories and by subsequently integrating the SFRs across cosmic time. In addition, the stellar mass accreted by mergers with other galaxies is added from the infalling satellites with stellar mass initialised at infall, as  described in Sect. \ref{sec_method_grow_Mstar}.

    \begin{figure}
        \includegraphics[width=0.99\columnwidth]{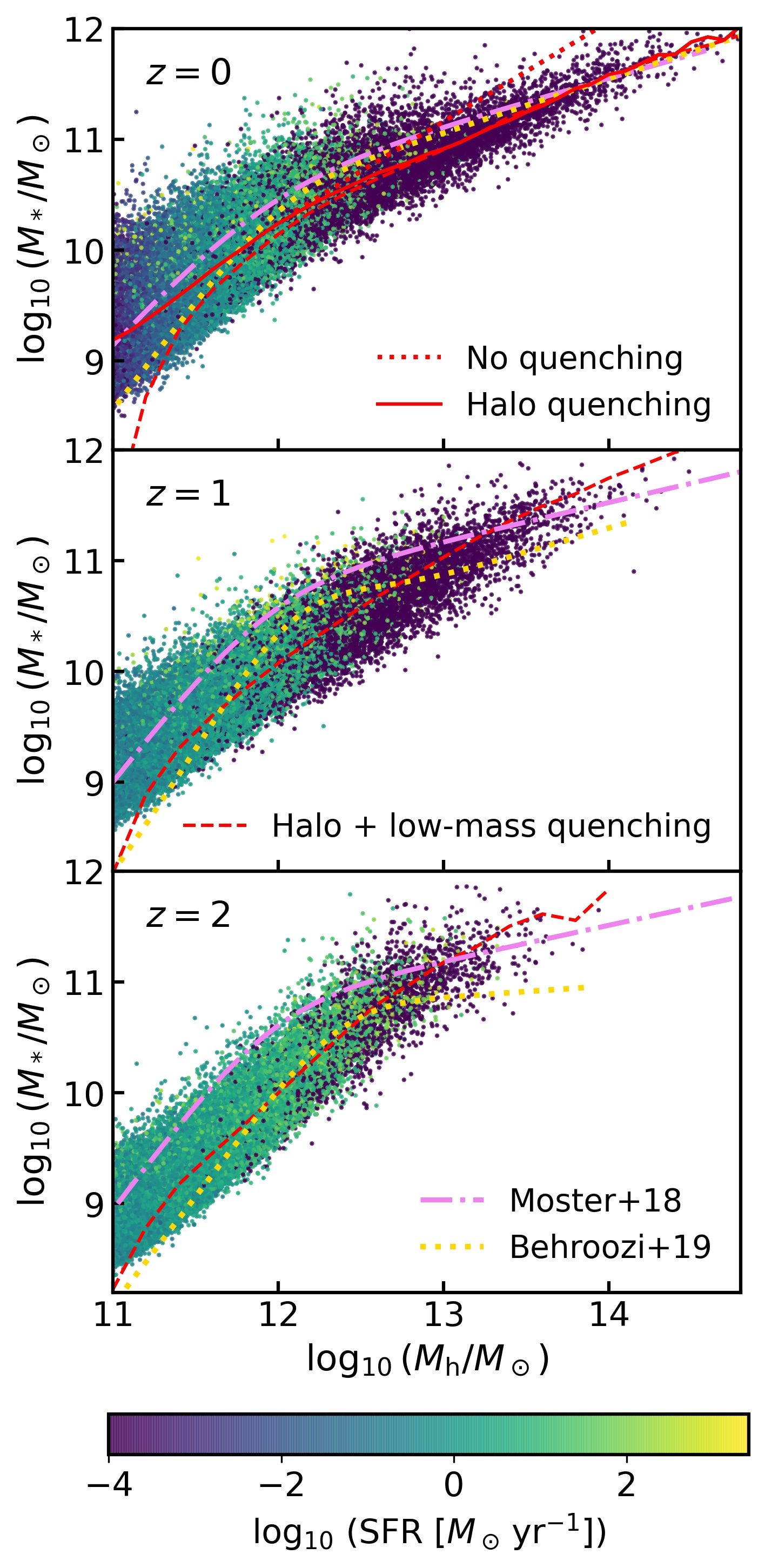}
        \caption{Distribution of the galaxies in \decode's catalogue on the $M_\star-M_{\rm h}$ plane at redshifts $z=0, 1 \,{\rm and}, \, 2$, for the halo quenching scenario. The colour bar represents the star formation rate of the galaxies. 
        The red dotted, solid and dashed lines show the scenarios with no quenching, halo quenching-only, and both halo and low-mass quenching, respectively. We also show the stellar mass-halo mass relations from \citet{moster_2018} (violet dash-dotted lines) and \citet{behroozi_2019} (yellow dotted lines), at the same redshifts.}
        \label{fg_SMHM_haloquenching}
    \end{figure}
    
    Figure \ref{fg_mass_acc} shows an example of evolution of a quenched galaxy and its host dark matter halo in \decode's catalogue. The galaxy initially forms stars at a rate given by the SFR-HAR relation following the dark matter HAR evolution. Then, at $z\sim0.9$, when the halo reaches the critical mass of quenching, the SFR drops instantaneously and the galaxy is labelled as quenched in the catalogue. After being quenched, the galaxy can still continue to grow in stellar mass ex-situ via mergers.

    \subsection{The stellar mass-halo mass relation and stellar mass function}\label{sec_res_SMHM_SMF}

    \begin{figure}
        \centering
        \includegraphics[width=\columnwidth]{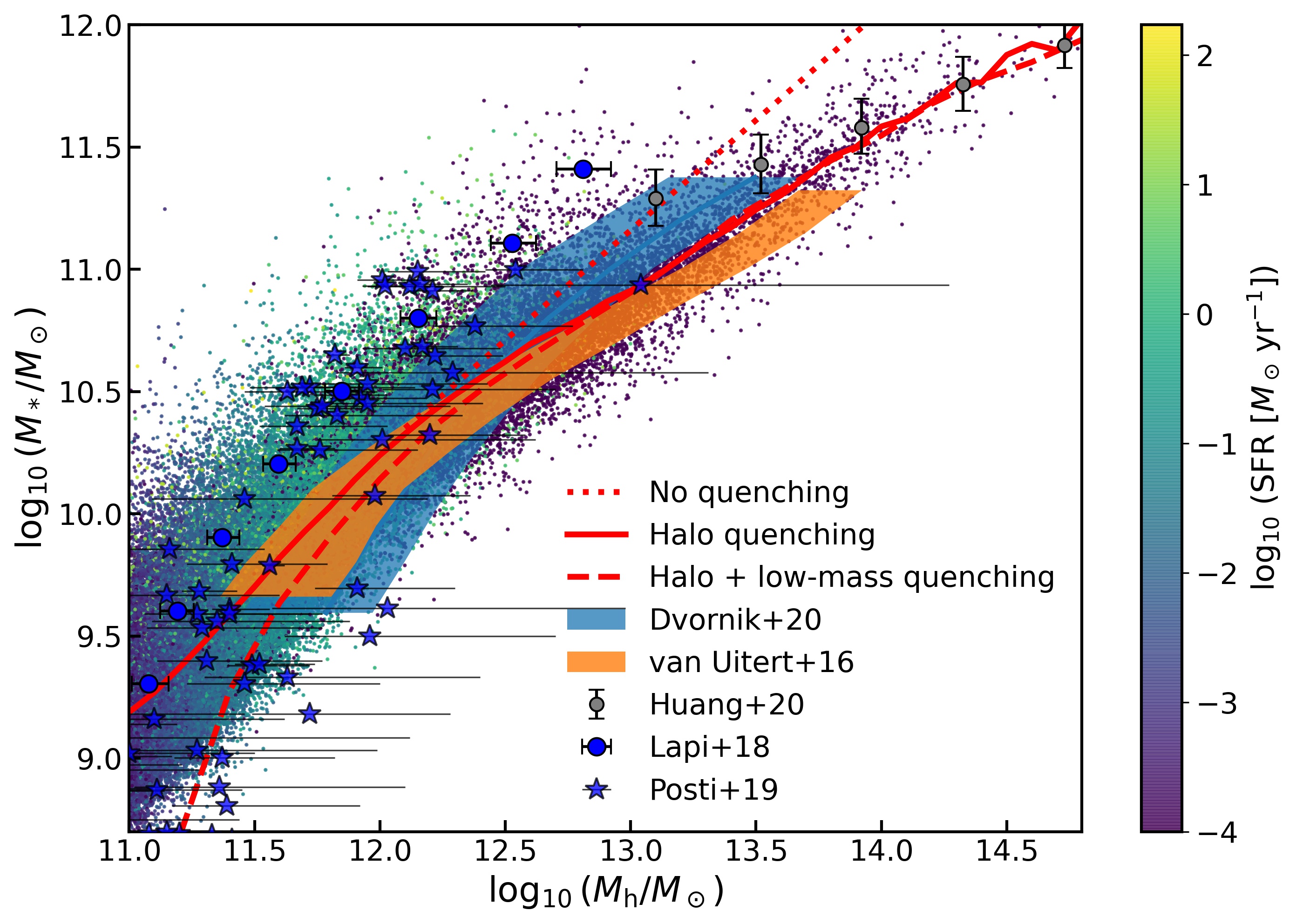}\\
        \includegraphics[width=0.9\columnwidth]{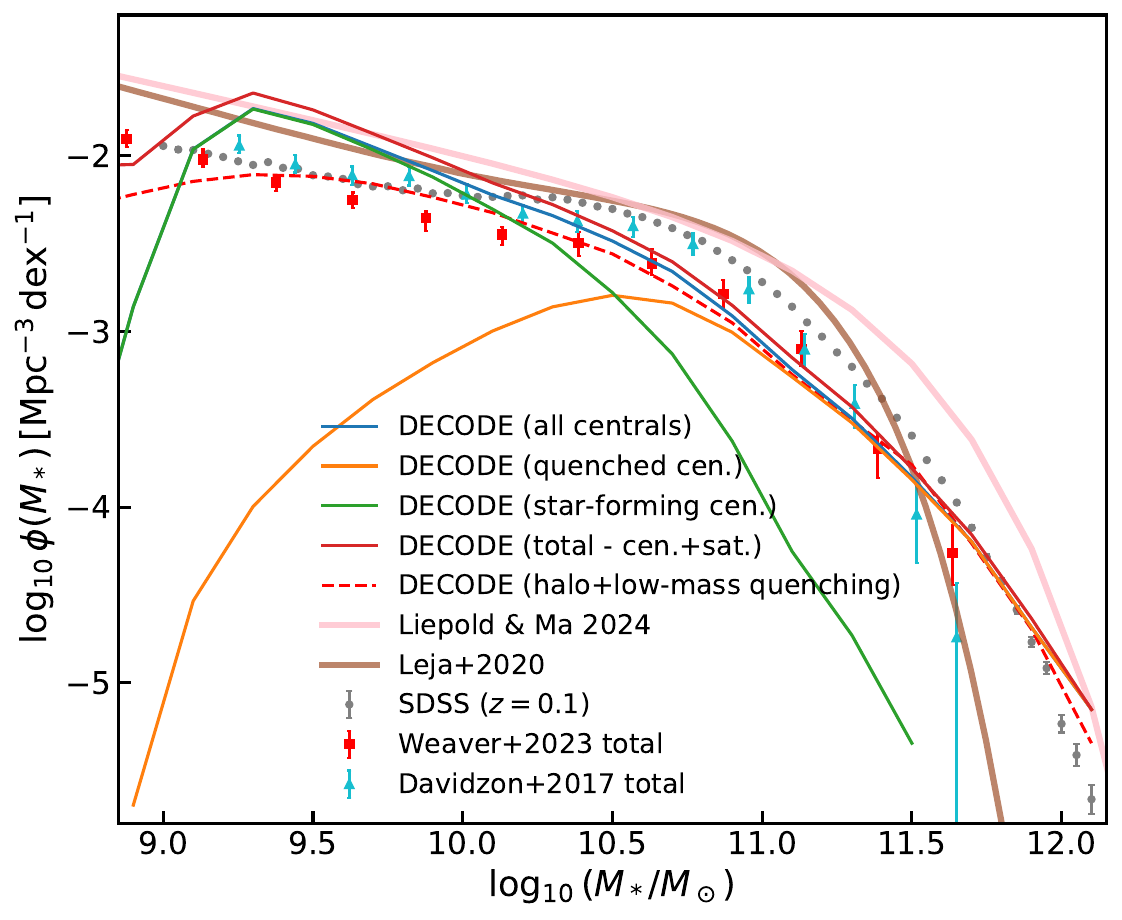}
        \caption{Upper panel: Same as Fig. \ref{fg_SMHM_haloquenching} at redshift $z=0$ with comparison to observations. The predictions of \decode are compared to the stellar mass-halo mass relation from weak lensing determinations at low redshifts from \citet{van_uitert_2016} (orange shaded area), \citet{dvornik_2020} (blue shaded area) and \citet{huang_2020} (grey dots with error bars), as well as the local spiral galaxies data from \citet{lapi_2018_scalingrelations} (blue dots with error bars) and \citet{posti_2019} (blue stars with error bars). Lower panel: Stellar mass function predicted by \decode at redshift $z=0$. The blue, orange, and green lines show the stellar mass function for all galaxies, quenched galaxies and star-forming central galaxies. The red solid lines show the total stellar mass function inclusive of the population of satellite galaxies. The red dashed lines show the case where low-mass quenching is included (Sect. \ref{sec_low_mass_end}). The brown solid line shows the 3D-HST+COSMOS data (\citealt{leja_2020}) and the pink solid line shows the determination from MASSIVE (\citealt{liepold_2024}). The grey dots, red squares and cyan triangles with error bars show the observational total stellar mass functions from the SDSS at $z=0.1$ (\citealt{bernardi_2017}), COSMOS2020 (\citealt{weaver_2023}) and COSMOS2015 (\citealt{davidzon_2017}) surveys.}
        \label{fg_SMHM_SMF_comp_data}
    \end{figure}

    Equipped with the stellar mass assembly tracks of the galaxies along with the host dark matter accretion histories, a first quantity that \decode can naturally predict is the SMHM relation at any redshift. We stress again that the SMHM relation and its scatter are both predictions of \decode here, and not used as inputs as, e.g. in \citet{grylls_2020a} or \citet{fu_2022}. Similarly, also the statistical abundances of galaxies, described by the SMF is an output of the model. 

    Figure \ref{fg_SMHM_haloquenching} shows how the galaxies in \decode's mock catalogue are distributed on the stellar mass-halo mass plane at different redshifts ($z=0$, $1$, and $2$) depending on their SFRs. We also show, for comparison, the SMHM relations from different quenching scenarios (red lines) which we discuss below, as well as those from the semi-empirical models EMERGE (\citealt{moster_2018}) and UniverseMachine (\citealt{behroozi_2019}). The SFRs of galaxies are shown via the colour bar. We see that at high redshifts most galaxies are low mass and are forming stars at higher rates, due to the higher accretion rate of their dark matter haloes. When moving towards lower redshifts, more massive galaxies are being formed and the number of quenched galaxies increases gradually, as the accretion rate of the cold gas available for star formation gradually drops, a phenomenon known as cosmological starvation (\citealt{feldmann_2015}). Furthermore, in the upper panel of Fig. \ref{fg_SMHM_SMF_comp_data} we compare our predictions to the SMHM relations inferred by weak lensing determinations from the KiDS and GAMA surveys (\citealt{van_uitert_2016} and \citealt{dvornik_2020}) and the Hyper Suprime-Cam survey (\citealt{huang_2020}), along with the relation of blue spirals from \citet{lapi_2018_scalingrelations} and \citet{posti_2019}.

    The lower panel of Fig. \ref{fg_SMHM_SMF_comp_data} shows the SMF for our galaxies at redshift $z=0$. We compare \decode's predictions with the latest data sets on the SMF at the same redshifts from COSMOS2015 (cyan triangles with error bars; \citealt{davidzon_2017}), COSMOS2020 (red squares with error bars; \citealt{weaver_2023}), 3D-HST+COSMOS (brown solid lines; \citealt{leja_2020}), as well as those in the local Universe of MASSIVE (pink solid line; \citealt{liepold_2024}) and SDSS (grey dots with error bars; \citealt{bernardi_2017}). The blue, orange and green solid lines show the SMF of all galaxies, quiescent galaxies, and star-forming central galaxies in the catalogue, respectively, and the red solid lines show the SMF of all galaxies including the addition of satellite galaxies (see Sect. \ref{sec_method_sat_gals}). We comment on the results in more detail below.

    \subsubsection{The low-mass end}\label{sec_low_mass_end}


    In the first instance, when assuming no quenching at all, the integration of the SFRs implied by the input SFR-HAR relation produces SMFs that are close in shape to the single Schechter halo mass functions, but simply rescaled by the mean ratio between SFR and HAR, as expected. It is interesting to note that this basic recipe yields a local SMF consistent with the one recently calibrated by \citet{leja_2020} and \citet{liepold_2024}, at least in the low-mass range at $M_\star \lesssim 3 \times 10^{10} \, M_\odot$, but overestimating any other determination of the local SMF (lower panel of Fig. \ref{fg_SMHM_SMF_comp_data}). The same no-quenching model also yields a single power law correlation between stellar mass and halo mass for the star-forming galaxies\footnote{The star-forming galaxies follow a steeper relation with respect to the mean overall relation.} which is broadly consistent with the direct estimates of the SMHM relation in star-forming spirals in the local Universe performed by \citet{lapi_2018_scalingrelations} and \citet{posti_2019}, as shown in the upper panel of Fig. \ref{fg_SMHM_SMF_comp_data}, further supporting the scenario in which massive local spirals follow a SMHM relation steeper than that of quenched galaxies. Even after switching on halo quenching, several of our massive galaxies, namely those residing at the centre of lower mass haloes with $M_\star \lesssim 5\times 10^{11} \, M_\odot$, avoid the quenching and conserve a high SMHM ratio, in agreement with the dynamical data. In the present framework therefore, these local massive spirals would be consistent with an evolution clear of mergers but sustained by continuous gas cooling, ultimately linked to the rate of mass accretion onto the parent halo, which would be in line with the theory of conservation of angular momentum between haloes and galaxies (e.g. \citealt{zanisi_2020}). 
    

    Furthermore, we have also checked that the contribution of merging satellites does not appreciably affect the total stellar mass growth of the less massive central galaxies (with present-day stellar mass below $M_\star \lesssim 10^{11} \, M_\odot$), and therefore the slope of the low-mass end of the SMHM relation and SMF, suggesting the need for additional physics to reproduce the more steeper slope of the SMHM relation and flatter slope of the double-Schechter SMF below the knee ($M_\star \lesssim 10^{11}\, M_\odot$). The excess at the low-mass end of \decode's SMF with respect to the observational SMF could be due to the inefficiency of the halo quenching process, suggesting that the action of other types of quenching mechanisms halts star formation in less massive galaxies. As shown by the red dashed lines in the lower panel of Fig. \ref{fg_SMHM_SMF_comp_data}, if we include the energy release, for example from strong supernova (SN) feedback that heats the gas, by simply shutting down the star formation in haloes with mass $\lesssim 10^{11}\, M_\odot$ (hereafter referred to as low-mass quenching) following what was done in several analytical models in the literature \citep[e.g.][]{springel_2001, springel_2003, cirasuolo_2005, shankar_2006, cattaneo_2008, governato_2010, brook_2011, pontzen_2017, henriques_2019}, the resulting SMF is much more compatible with observations and also the slope of the SMHM relation's faint end is much steeper, supporting the need for additional processes at low stellar masses.

    \subsubsection{The high-mass end}
    
    We now move our attention to the high-mass end of the SMHM relation and SMF. On the assumption of continuous star formation, the slope of the SMHM relation would be much steeper with respect to other predictions and weak-lensing determinations of early-type galaxies for example. Also the slope of the SMF's bright end would be much flatter, being heavily overestimated with respect to the observational measurements. By activating the halo quenching mechanism with the threshold mass at $M_{\rm h} \sim 10^{12}\, M_\odot$, above which galaxies are mostly all quenched, the SMF's high-mass end is steepened due to the halting of star formation in more massive galaxies, leading also to a corresponding flattening in the SMHM relation, in better agreement also with other semi-empirical models, 
    such as EMERGE and UniverseMachine, especially at $z=0$, and with other independent estimates of the SMHM relation from weak lensing in the local Universe. 
    We also find that the scatter in stellar mass at fixed halo mass increases with redshift and decreases with halo mass, as also suggested by other works (e.g. \citealt{matthee_2017, allen_2019, erfanianfar_2019}), but without any apparent dependence on SFR at fixed halo mass. In other words, the dispersion in the scatter in this approach is mostly induced by the varied host halo mass assembly tracks, and not by variations in SFRs at fixed halo mass. These findings do not change by altering the scatter in SFR at fixed HAR within reasonable values, as anticipated in Sect. \ref{sec_method_AM_sfr_har}.

    \begin{figure*}
        \includegraphics[width=\textwidth]{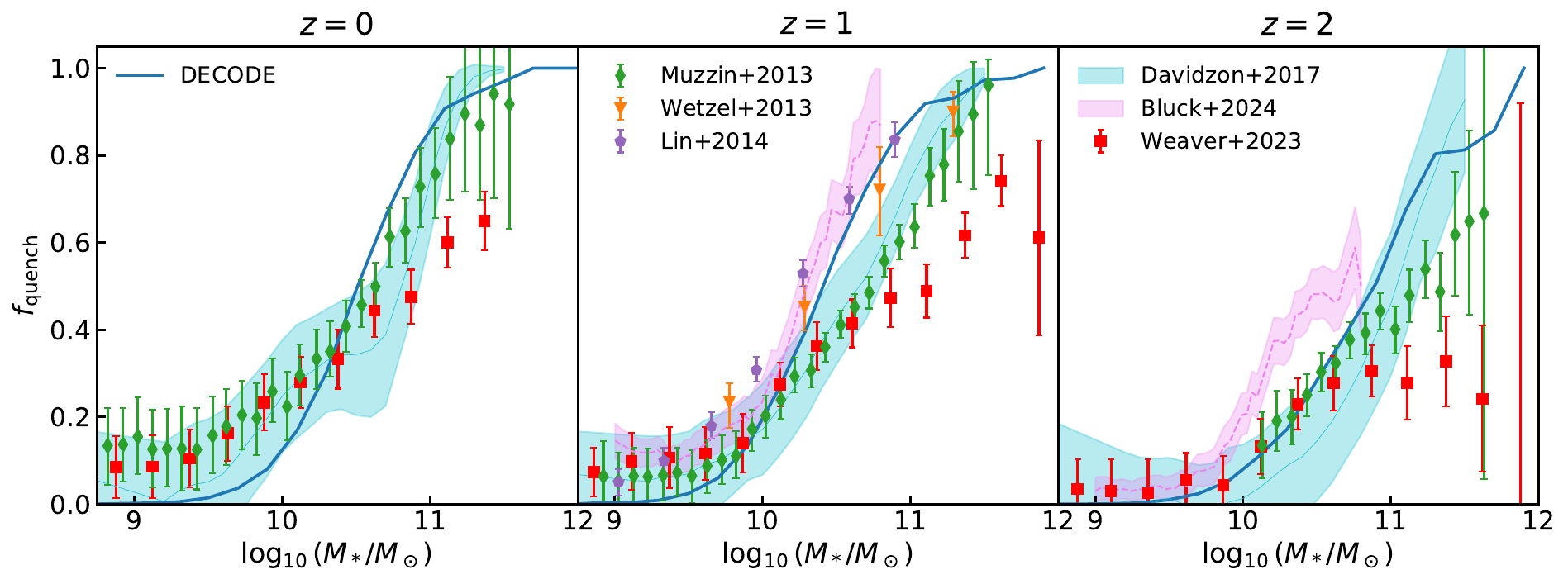}
        \caption{Fraction of quenched galaxies as a function of stellar mass at redshifts in the same redshift bins. The predictions from \decode are shown for central galaxies only (blue solid lines). We compare \decode's predictions with the observed quenched fractions from the COSMOS2015 (cyan solid lines and shaded areas; \citealt{davidzon_2017}) and COSMOS2020 (red squares with error bars; \citealt{weaver_2023}) surveys. Further data from \citet{muzzin_2013} (green rhombuses with error bars), \citet{wetzel_2013} (orange triangles with error bars), \citet{lin_2014} (purple pentagons with error bars), and \citet{bluck_2024_quenchJWST} (JWST-CEERS at $1<z<2$ and $2<z<4$, violet dashed lines and shaded areas) are shown for comparison.}
        \label{fg_fquenched}
    \end{figure*}

    Furthermore, as \decode is currently implemented, the halo quenching scenario predicts a SMF broadly in agreement with the COSMOS2015 and COSMOS2020 observed SMFs, and with SDSS, in the local Universe. In particular, the high-mass end of the SMF produced by \decode is in relatively good agreement with the observed SMFs at $z\sim0$, showing the efficacy of this quenching mechanism at low redshift. 

    \subsection{The fraction of quenched galaxies}\label{sec_res_SMF_fq}

    We show in the previous Sections that, all in all, a proportionality between SFR and HAR is sufficient to generate a SMF with a Schechter shape broadly aligned with the data in the local Universe. Nevertheless, this approach tends to somewhat underproduce the number density of galaxies around the knee, especially with respect to the SDSS SMF by \citet{bernardi_2017}. This discrepancy could be ascribed to an underlying inconsistency between the determinations of the SFRs and the estimates of stellar masses in the SMF. We have verified that when using the SFR-HAR relation from the TNG, with SFRs self-consistently linked to the stellar masses in the simulation, the resulting SMF is in better agreement with the SDSS data. 

    Irrespective of the quality of the match with the total SMF, the predicted relative fractions of quenched galaxies as a function of stellar mass or environment could be considered an independent and robust test of the quenching mechanism. Figure \ref{fg_fquenched} shows the fraction of quenched (central) galaxies as a function of stellar mass predicted by \decode at redshifts $z=0$, $1$ and $2$, compared to those observationally inferred from the COSMOS/UltraVISTA (green rhombuses; \citealt{muzzin_2013}), SDSS (orange triangles; \citealt{wetzel_2013}), Pan-STARRS1 Medium-Deep (purple pentagons; \citealt{lin_2014}), COSMOS2015 (cyan shaded areas; \citealt{davidzon_2017}), COSMOS2020 (red squares; \citealt{weaver_2023}) and JWST-CEERS (violet shaded areas; \citealt{bluck_2024_quenchJWST}) surveys. The general behaviour of the quenched fraction is broadly reproduced. We note that towards higher stellar masses above the halo mass threshold, all galaxies are quenched, agreeing overall with the quiescent fraction from observations, while towards low stellar masses all galaxies in our catalogue are star-forming, as they live in dark matter haloes below the critical mass for quenching. We also note that the latest data from JWST-CEERS from \citet{bluck_2024_quenchJWST} point to somewhat larger fractions of quenched galaxies at $z>2$ with respect to other data and also to what \decode can predict. This slight but significant discrepancy may be induced by either differences in stellar mass and SFR estimates between data and the input SFR functions from \citet{mancuso_2016}, or could indicate the action of alternative and/or additional quenching mechanisms acting at high masses and/or at higher redshifts. We will further study the very high redshift Universe from the latest data from \decode in a separate dedicated paper where we will test the effectiveness of different quenching mechanisms against the latest high-z JWST data.

    \begin{figure}
        \includegraphics[width=\hsize]{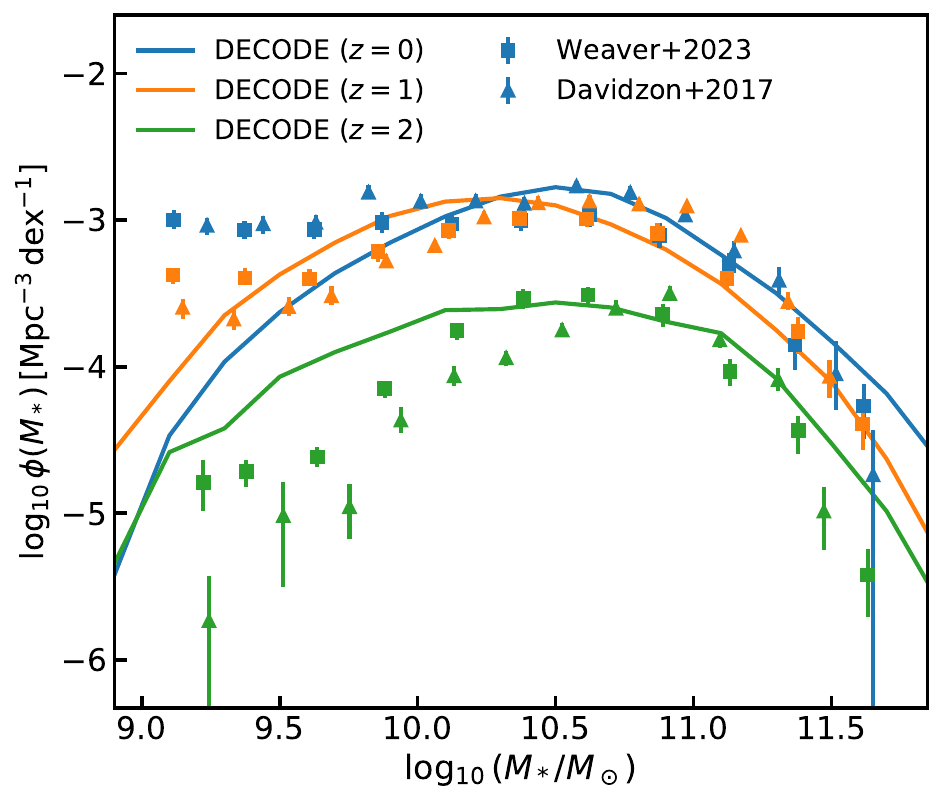}
        \caption{Stellar mass function of quenched galaxies as predicted by \decode (coloured solid lines) at redshifts $z=0$, $1$, and $2$, compared to those inferred by COSMOS2015 (\citealt{davidzon_2017}, coloured triangles with error bars) and COSMOS2020 (\citealt{weaver_2023}, coloured squares with error bars).}
        \label{fg_SMFquenched}
    \end{figure}

    Finally, in Fig. \ref{fg_SMFquenched}, we show the SMF of quenched galaxies, compared to the observationally determined SMFs from \citet{davidzon_2017} and \citet{weaver_2023}. A comparison with the SMFs of quenched galaxies better highlights any shortfall of the quenching model. 
    The comparison shows that there is good agreement between \decode's prediction and the observed SMF above $M_\star \gtrsim 10^{10} \, M_\odot$, proving the efficiency of the halo quenching in massive galaxies. The disagreement at redshifts $z=0$ and $1$ below $M_\star \lesssim 10^{10} \, M_\odot$ could be associated with the need for additional quenching mechanisms (e.g. morphological quenching or feedback processes), or some environmental quenching for the satellite quenched population which could slightly affect the overall statistics. We will show the impact of these factors in the next papers of this series. Finally, the overestimation of \decode's quenched SMF with respect to the inferred SMFs is simply due to an excess of low-mass galaxies in the entire galaxy population in \decode.


\section{Discussion}\label{sec_discuss}

In this work, we put forward a new data-driven semi-empirical methodology and its proof-of-concept application to test the role of several physical processes, such as star formation and quenching, involved in the galaxy stellar mass assembly. We showed that this semi-empirical approach, based on a minimal number of assumptions and parameters, can be extremely useful to address such problems in a transparent way. 
In particular, \decode is a data-driven model that uses as input observational quantities, such as the SFR distribution, to build-up galaxy stellar mass across cosmic time. 


First of all we showed that the hypothesis of monotonicity between galaxy SFR and dark matter HAR, corroborated also by simulations recent observations (e.g. \citealt{daddi_2022, sillassen_2024}), produces galaxy abundances in agreement with the latest determinations of the SMF. This is also in agreement with the results from \citet{lapi_2017}, who showed an impressive agreement between the SFR function found by \citet{mancuso_2016} and the SMF, using a continuity equation approach. Here, we confirm that a similar broad agreement can be obtained even with the abundance matching between SFR and HAR and with the halo quenching mechanism to regulate the halting of star formation. This finding highlights the response of the gas cooling rate (i.e. SFR) to the infalling gas rate of the host dark matter halo (i.e. HAR), whereas in previous models it was mostly connected to the available gas amount (i.e. halo mass). However, in the current implementation of \decode, the galaxy SMF around the knee ($M_\star \sim 10^{11} \, M_\odot$), which is a key stellar mass range where quenching and morphological transition typically happen, is slightly underestimated with respect to observations. The reason behind this deviation is the following. Should one exclude quenching at all, the input observed SFRs are enough to produce enough low- and intermediate-mass galaxies, but heavily overestimate the bright end of the SMF, and produce a single Schechter shape, which is a simple reflection of the accretion histories of the dark matter encoded in the HMF. Therefore, a strong and instantaneous halo quenching could be not a realist scenario for modelling quenching processes in massive galaxies at low redshifts. 
Therefore, we tested a time-delayed quenching model (e.g. \citealt{wetzel_2013, haines_2013, haines_2015, lian_2016, foltz_2018, akins_2021, tacchella_2022, reeves_2023, bravo_2023}) according to which the SFR declines exponentially over $\sim 1-2$ Gyr instead of instantaneously which, however, yields resulting galaxy stellar mass assemblies and therefore the SMF that does not change appreciably. Additionally, state-of-the-art hydrodynamical simulations have shown that, e.g. AGN feedback has also an extremely important role in quenching massive ellipticals (e.g. \citealt{su_2021, wellons_2023}). We will incorporate the AGN physics in the model in the next paper of this series where we will investigate the role of each model and how their combination will better align with the observed galaxy abundances at each redshift. Lastly, in our model we assume the halo mass quenching threshold to be $10^{12} \, M_\odot$, as suggested by \citet{dekel_2006}, but its actual value and the dispersion around it can also sensibly alter the output stellar mass growths.


We proved that the halo quenching mechanism is efficient in halting star formation in massive galaxies ($M_\star \gtrsim 3 \times 10^{10} \, M_\odot$), being able to match the observed quiescent SMF and quenched fractions up to $z\sim2$. \decode is also able to reproduce the vast majority of the local determinations of the SMHM relation from weak lensing and spirals. 
We also found some dependence for the galaxy stellar mass on the formation age of the host dark matter halo (i.e. when it reaches half of the present-day mass), with more massive galaxies favouring older haloes and vice versa, also found in the recent work of \citet{oyarzun_2024} from the SDSS catalogue. In addition, at redshifts $z\simeq2$ \decode's predictions for the quenched abundances are slightly below the JWST data (\citealt{bluck_2024_quenchJWST}), as shown in Fig. \ref{fg_fquenched}, because the halo quenching mechanism is not efficient at $z\gtrsim 2$. This tiny deviation with the data may be induced by other quenching mechanisms at those redshifts, such as AGN feedback. The recent work of \citet{dekel_2023} showed that the high-redshift luminous galaxies may grow with feedback-free starburst and with their star formation being suppressed by feedback at later epochs. This work will be a pathfinder for our dedicated analysis on high-$z$ galaxies in the next papers of this series. 


Finally and interestingly, the very recent determinations from JWST (e.g. \citealt{nanayakkara_2022, carnall_2023, valentino_2023, dome_2024, weibel_2024}) showed a relatively large population of quenched galaxies even at higher redshifts ($z\gtrsim5$), in contrast to the predictions of numerical simulations \citep[see e.g.][]{vani_2024}. This unexpectedly large amount of observed quenched galaxies from JWST could further indicate the need for additional or simply different, more widespread and efficient quenching mechanisms. To disentangle this, a possible way to address quenching at high redshifts with \decode could be to start building up galaxies at a higher redshift, instead of $z=3$. The recent high-redshift ($z\gtrsim 8$) determination of the UV luminosity function (e.g. \citealt{donnan_2023, donnan_2024, harikane_2023}), combined with the available data at lower redshifts, will be a precious input in \decode to extend our investigation in the early Universe. With these inputs, \decode will shed light on the newly observed massive galaxies at high redshifts that were missing in the previous cosmic census and cosmological models.

\section{Conclusions}\label{sec_conclu}

In this paper, we have presented our new semi-empirical model, \decode, to study the stellar mass formation and evolution of galaxies. To this purpose, we made use of \decode in a $\Lambda$CDM framework and converted dark matter assembly into galaxy star formation histories via the input SFR-HAR relation. We started by testing \decode's assumption on the existence of a connection between the galaxy SFR and the host dark matter halo accretion rate. We found in the TNG that the galaxy SFR and the accretion rate of the host dark matter halo are connected to each other and can be modelled via the abundance matching technique. We then switched to real data and made our prediction using observationally determined SFRs as input.

The main results of this paper can be summarised as follows:

\begin{itemize}
    \item We showed the presence of a monotonic proportionality between the galaxy SFR and dark matter HAR, as also supported by cosmological hydrodynamical simulations like the TNG, which is relatively constant in the range $1<z<3$, and dropping in normalization only at $z<1$ (Fig. \ref{fg_SFR_HAR_TNG}).
    \item We showed that such a relation is capable of reproducing the bulk of the observed galaxy population and SMHM relation in the local Universe. 
    \item We showed that the scatter in stellar mass at fixed halo mass decreases with increasing halo mass, consistently with what is predicted by other models.
    \item We showed via \decode that the star formation activity in most massive galaxies is efficiently halted by the halo quenching process, at least at redshifts $z\lesssim 1-2$, which allows to significantly improve the alignment of the model predictions with the local SMF and SMHM relation at high stellar masses. 
    \item To produce the steep (flat) faint end of the SMHM relation (SMF) observed in some data sets, some quenching mechanism in lower mass haloes is needed. By setting a cut in halo mass, below $10^{11}\, M_\odot$ to mimic, for example, the effect of SN feedback or even the putative effect of AGN feedback in low mass galaxies (Sect. \ref{sec_low_mass_end}), we induced a steeper slope in the SMHM relation. However, some of the latest current data on the SMF do not require this additional ingredient, as they have steeper faint end slopes. 
\end{itemize}

In conclusion, the SFR-HAR mapping is a powerful new way of modelling galaxy evolution and star formation histories, along with some assumption on the quenching of star formation. It can accurately and rapidly reproduce galaxy stellar mass assemblies, stellar mass functions, relative amounts of star-forming and quiescent galaxies, and SMHM relations. \decode will also constitute a very valuable tool for studying galaxies at very high redshifts thanks to the advent of new high-quality data from the ongoing missions such as JWST \citep[][]{gardner_2006} and Euclid \citep[][]{amiaux_2012, euclid2024_overview}.

\begin{acknowledgements}
This work received partially funding from the European Union’s Horizon 2020 research and innovation programme under the Marie Sk\l odowska-Curie grant agreement No. 860744 for the BiD4BESt project (coordinator: F. Shankar). HF acknowledges support at Fudan University from the Shanghai Super Post-doctoral Excellence Program grant No. 2024008. L.B. acknowledges financial support from the German Excellence Strategy via the Heidelberg Cluster of Excellence (EXC 2181 - 390900948) STRUCTURES. MA is supported at the Argelander Institute f\"ur Astronomie through the Argelander Fellowship. Y.P. acknowledges support from the National Science Foundation of China (NSFC) grant Nos. 12125301, 12192220 and 12192222, the science research grants from the China Manned Space Project with No. CMS-CSST-2021-A07, and support from the New Cornerstone Science Foundation through the XPLORER PRIZE. F.Y. was supported in part by NSFC (12133008, 12192220, and 12192223).
\end{acknowledgements}

\bibliographystyle{aa}
\bibliography{A&A_v3/main}


\appendix



\section{Additional star formation rate functions}\label{Appendix:additional_SFRs}

\begin{figure}
\centering
    \includegraphics[width=0.9\hsize]{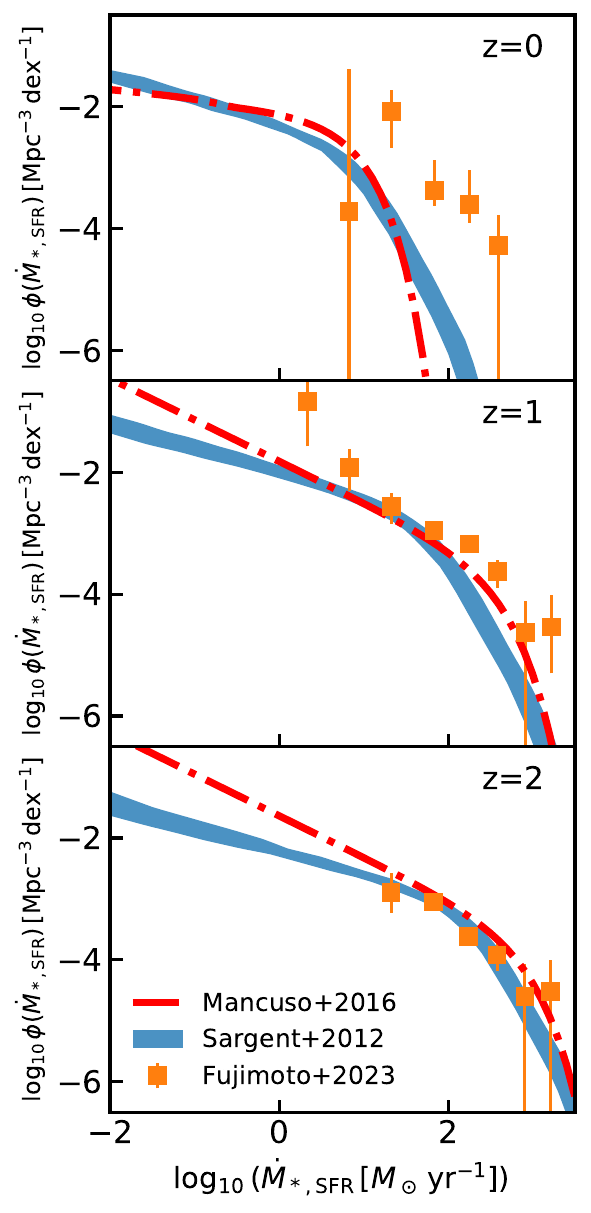}
    \caption{Star formation rate function from \citet[][red dash-dotted lines]{mancuso_2016}, \citet[][blue shaded areas]{sargent_2012}, and \citet[][orange squares with error bars]{fujimoto_2023} at different redshifts.}
    \label{fg_phiSFR_comparison}
\end{figure}

In this Appendix, we comment how the results on the galaxy stellar mass growths would change by employing different data sets for the SFR function. Figure \ref{fg_phiSFR_comparison} shows the SFR function from \citet{sargent_2012} and \citet{fujimoto_2023}, in addition to that from \citet{mancuso_2016} already presented in Sect. \ref{sec_sfr_har_relation}. 
The \citet{fujimoto_2023} SFR function is much steeper at the faint end compared to the others at all redshifts and at the bright end contains a higher number of bright objects. Instead, the SFR function of \citet{sargent_2012} has a flatter faint end with respect to the others at $z>0$. The corresponding SFR-HAR is consequently characterised by a flatter slope for a steeper SFR function and vice versa.

The flatter SFR-HAR relation from the \citet{fujimoto_2023} inputs produces a SMF with a high excess of low-mass galaxies due to the high SFR, and results into lower quenched fractions. On the other hand, the \citet{sargent_2012} input SFR function predicts opposite results.

\section{Major merger rates}\label{sec_merger_rates}

\begin{figure}
\centering
    \includegraphics[width=0.95\hsize]{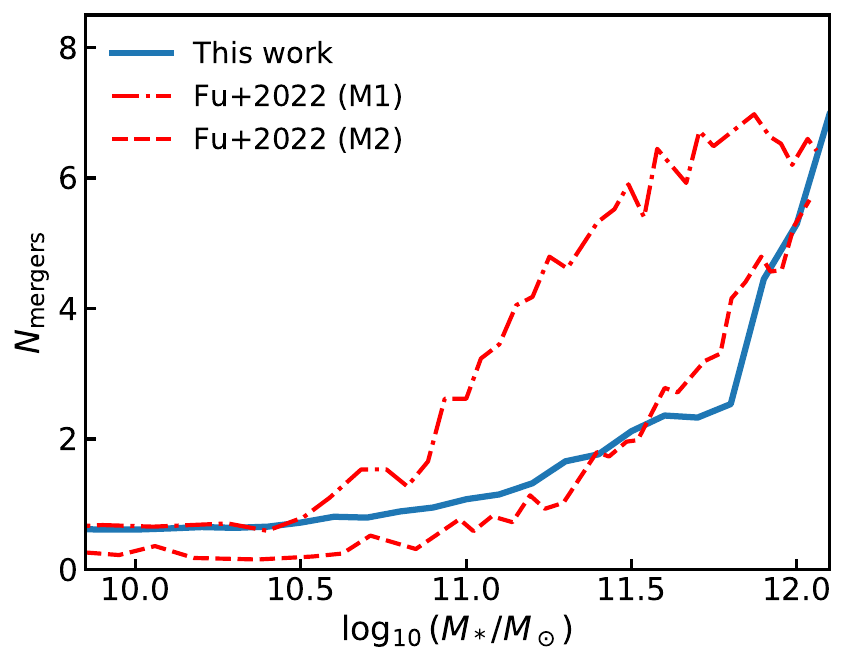}
    \caption{Number of galaxy major mergers, with mass ratio $ M_{\rm \star,sat} / M_{\rm \star,cen} > 0.25$, as a function of the central galaxy mass from \decode (blue solid line) and the \citealt{fu_2022} Models 1 and 2 (red dash-dotted and dashed lines, respectively).}
    \label{fg_N_major_mergers}
\end{figure}

As an additional application of \decode, we show its ability in predicting the galaxy merger rates. Mergers are an extremely important factor in the galaxy evolution and, in addition to the stellar mass assembly, they also play a crucial role in shaping galaxy morphologies and bulge formation (e.g. \citealt{cole_2000, hatton_2003, bournaud_2007, hopkins_2010b, guo_2011, shankar_2013, fontanot_2015, croton_2016, cattaneo_2017}). In Fig. \ref{fg_N_major_mergers}, we show the average number of major mergers as a function of the final galaxy stellar mass at $z=0$. The plot shows that \decode also predicts an increasing number of mergers as a function of stellar mass, similarly to what was found in \citet{fu_2022}, which is due to the broken power law shape of the SMHM relation. We stress that the presence of major mergers due to the flatting of the SMHM relation, without which they would not take place, is a direct product of the quenching in \decode. Such shape of the merger rates also allows to predict a comparable fraction of elliptical galaxies and bulge-to-total ratio distributions, by adopting a toy major merger model as in \citet{fu_2022}.

Finally, interestingly, the mergers implemented following the recipe described in Sect. \ref{sec_method_grow_Mstar} and shown in Fig. \ref{fg_N_major_mergers} are characterised by slightly slower increasing rate and mostly constant rate for the less massive galaxies, unlike what was found by previous works (e.g. \citealt{hopkins_2010, grylls_2020b, fu_2022}), even though the overall trend is consistent. This is due to the shape of the SMHM relation, which does not present a remarkable bending between the low-mass and high-mass ranges. Instead, we tested that by using a visible double power law-shaped relation the output number of major mergers would be increasing as a function of stellar mass, further supporting the findings that we presented in \citet{fu_2022}.

\end{document}